\documentclass[12 pt]{article}
  
\usepackage{ucs} 
\usepackage[utf8x]{inputenc}   
\title{\LaTeX}  
\date{}  
\author{}

\usepackage{amsmath, amsfonts, amssymb}
\usepackage{graphicx}
\usepackage{indentfirst}
\usepackage{threeparttable}
\usepackage{url}

\def\la{\;
\raise0.3ex\hbox{$<$\kern-0.75em\raise-1.1ex\hbox{$\sim$}}\; }
\def\ga{\;
\raise0.3ex\hbox{$>$\kern-0.75em\raise-1.1ex\hbox{$\sim$}}\; }

\newcommand{\kms}{km~s$^{-1}$}

\newcommand{\dmm}{$\Delta\mu/\mu$}

\begin{document}

\title{\bf Constraints on $\mu$-variations from the 
methanol CH$_3$OH thermal lines in the Galaxy at
galactocentric distances of $1.5 < D_{GC} < 13$ kpc}
\author{\bf A. I. Shtein$^{1}$, 
J. S. Vorotyntseva$^{1,2*}$, 
S. A. Levshakov$^{2}$}
\maketitle
\date{\it  \small  $^1$Department of Physics, St. Petersburg
 Electrotechnical University ``LETI'', Prof. Popov Street 5, 
197376, St. Petersburg, Russia\\
$^2$Ioffe Institute, Politekhnicheskaya Street 26, 194021
St. Petersburg, Russia}\\

\thanks{$^*$E-mail: j.s.vorotyntseva@mail.ioffe.ru}

\begin{abstract} 
In this study, constraints are set on hypothetical variations 
of the fundamental physical constant 
$\mu = m_{\rm e}/m_{\rm p}$~-- the electron-to-proton 
mass ratio~-- based on thermal emission lines of 
$E$-methanol (CH$_3$OH) in the 216--220 GHz frequency range.
Molecular spectra toward seven massive clouds located across 
a wide range of galactocentric distances ($1.5 < D_{GC} < 13$ kpc) 
have been analyzed.
The obtained weighted mean value
$\langle \Delta\mu/\mu \rangle= (0.5\pm0.5)\times10^{-7}$
does not indicate statistically
significant changes in $\mu$ at 
the $1\sigma$ confidence level of $5\times10^{-8}$, which is in good agreement with previously
established upper limits in a number of molecular
clouds in the Galactic disk.
\end{abstract} 

%{\it Key words}: masers~-- methods: observational~-- techniques: spectroscopic~-- ISM:
%molecules~-- elementary particles.

%----------------------------Section-1
\section{Introduction}
\label{Sect-1}

The search for hypothetical variations of fundamental physical 
constants is one of the most attractive tasks in modern laboratory 
and astrophysical studies.
The relevance of this topic stems from the recent active discussions 
of theories extending the Standard Model of particle physics~-- 
specifically, theories concerning the so-called dark sector, 
comprising dark matter (DM) and dark energy (DE). 
Among these, the most promising are considered to be theoretical 
models of dark matter in the form of scalar fields that can interact 
with elementary particles via the Higgs scalar field: 
directly with the electron, and with the proton via quarks. 
 Such interactions would, in turn, lead to a modulation of 
 the masses of these particles and, consequently, to a change 
 in one of the fundamental physical constants~--
 $\mu = m_{\rm e}/m_{\rm p}$, the electron-to-proton mass ratio 
 (see the detailed review \cite{Uzan}).
  
In astrophysical applications, the constancy of $\mu$ can be tested, for example, by observing molecular spectra. 
In 1993, it was shown that the 
electro-vibro-rotational transitions in the H$_2$ molecule exhibit 
a specific dependence on $\mu$ that is unique to each 
transition \cite{VL93}:
 \begin{equation}
 \tilde{f} = f + q\frac{\Delta\mu}{\mu}\, ,
\label{E1}
\end{equation}
where $\Delta\mu/\mu = (\mu_{obs} - \mu_{lab})/\mu_{lab}$ 
is the relative difference between the laboratory value $\mu_{lab}$ 
and the value in the astrophysical object $\mu_{obs}$, 
$f$ is the laboratory transition frequency, and $\tilde{f}$ 
is the shifted frequency in the comoving reference frame  
when $\Delta\mu/\mu \neq 0$.
In this expression, the $q$-factor is calculated using quantum-mechanical methods and is defined as the difference $q = q_u - q_\ell$ between the $q$-factors of the upper ($u$) and lower ($\ell$) energy levels.
 
 Dividing both sides of (\ref{E1}) by the laboratory frequency, 
we obtain the dimensionless sensitivity coefficient 
$Q = q/f$, which is used to derive an expression for estimating 
\dmm\ by comparing the radial velocities 
$V_i$ and $V_j$ of molecular lines with different coefficients 
$Q_i$ and $Q_j$ \cite{LKR}:
 \begin{equation}
 \frac{\Delta\mu}{\mu} = \frac{V_i - V_j}{c(Q_j - Q_i)}\, ,
\label{E3}
\end{equation}
where $c$ is the speed of light;
the radial velocity $V$ is calculated
from the measured astronomical frequency $\hat{f}$
in accordance with the radio-astronomical convention:
 \begin{equation}
\frac{V}{c} = 1 - \frac{\hat{f}}{f}\, .
\label{E4}
\end{equation}

Theoretical calculations have shown that methanol (CH$_3$OH) and its isotopologues~--
which are widespread in the molecular clouds of the Galactic disk~--
are the most sensitive molecules to changes in $\mu$.
For instance, for torsional-rotational transitions in methanol, 
the $Q$ values ​​lie in the range 
from $-53$ to $+42$ \cite{LKR, J11}; 
for $^{13}$CH$_3$OH, from $-32$ to $+78$; 
and for CH$_3$$^{18}$OH, the sensitivity coefficients 
vary within the range 
from $-109$ to $+33$ \cite{VKL24}.
Among the deuterated methanol isotopologues,
 CD$_3$OH stands out with $-300 < Q < +73$; 
 the other two deuterated molecules are comparable in sensitivity
to the parent methanol: for CH$_3$OD, the $Q$ coefficients range 
from $-32$ to +25, and for CD$_3$OD, from $-44$ to +38 \cite{VLK24}.

Other molecules used in $\Delta\mu/\mu$ estimates
have, for comparison, the following $Q$ values:
lines of the H$_2$ Lyman and Werner bands
are characterized by small sensitivity coefficients
$|Q| \sim 0.01$ \cite{VL93, Pot95, M06, Wim},
inversion transitions in ammonia
NH$_3$ have $Q = +4.46$ \cite{FK07},
and $Q$ coefficients for torsional-rotational transitions
in acetaldehyde CH$_3$CHO range
from +0.62 to +3.61 \cite{VLK26}.

In this work, our \dmm\ estimates are compared with data on sources in the Galactic disk located at various galactocentric distances $D_{GC}$.
Previous studies of individual molecular clouds have shown that 
modern radio-astronomical observations make it possible to achieve 
an accuracy in \dmm\ estimates on the level of $\sim 10^{-8}$.
 For instance, 
based on thermal CH$_3$OH lines\footnote{The term “thermal line” is used
in this paper to emphasize the distinction from purely maser
emission.}
 toward the molecular 
cloud L1498 ($D_{GC} \sim 8$ kpc), 
an upper limit of $|\Delta\mu/\mu| < 3 \times 10^{-8}$
was obtained \cite{Dap}\footnote{All estimates of physical quantities 
are given at the $1\sigma$ significance level.},
while toward Orion-KL ($D_{GC} \sim 9$ kpc), 
the limit is $|\Delta\mu/\mu| < 8 \times 10^{-8}$ \cite{VL25}.
Combinations of CH$_3$OH and CH$_3$CHO transitions in the molecular clouds
L1544, Barnard-1, and IRAS4A ($D_{GC} \sim 7–8$ kpc)
constrain the variation of $\mu$ at the level of
$|\Delta\mu/\mu| < 4 \times 10^{-8}$ \cite{VLK26}.
Finally, inversion transitions in ammonia, compared with 
purely rotational ($Q_{rot} = 1$) transitions in HC$_3$N, HC$_5$N, 
and HC$_7$N toward L1512 and L1498 ($D_{GC} \sim 8$ kpc), 
correspond to $|\Delta\mu/\mu| < 0.7 \times 10^{-8}$ \cite{L14}. 
Similar studies have also utilized methanol maser lines,
establishing an upper limit on $|\Delta\mu/\mu| < (2-3) \times 10^{-8}$ 
for a number of methanol masers at galactocentric distances
of $4 < D_{GC} < 12$ kpc \cite{L22, Ell}.

It is interesting to note that indications of statistically significant 
variations in $\mu$ were recently found near the Galactic center
toward the Sgr\,B2 molecular cloud ($D_{GC} \simeq 0.1$ kpc):
$\Delta\mu/\mu = (-3.4 \pm 0.4) \times 10^{-7}$ \cite{VL25, VLH, VL26}.
In this regard, continuing research into both the region near the Galactic center ($D_{GC} < 3$ kpc)~-- which remains poorly studied~--
and the Galactic periphery ($D_{GC} > 12$ kpc) remains a pressing task.
As is well known, these two regions in galaxies differ in the distribution 
of dark matter density, $\rho_{DM}$.
For example, the NFW profile \cite{NFW}, widely used in numerical calculations, predicts an increase in DM density near the center, 
$\rho_{DM} \propto r^{-1}$, and a decrease in the outskirts, 
$\rho_{DM} \propto r^{-3}$.
 
In this paper, we present estimates of $\Delta\mu/\mu$ obtained from thermal methanol lines observed in a number of sources spanning 
a range of galactocentric distances from 1.5 kpc to 13 kpc.

%-------------------------------Section 2
\section{Observation parameters and selection of objects for \dmm\ estimates}
\label{Sect-2}

Data from \cite{Zhao} were used for the estimates of $\Delta\mu/\mu$.
Observations were conducted using the 
IRAM 30-m telescope\footnote{The Institute for Radio Astronomy 
in the Millimeter 
Range  (IRAM) is an international research institute and Europe's
leading center for radio astronomy.}
in various spectral bands:
3 mm (90.6--98.2 GHz); 2 mm (108.0--115.4 GHz and 138.4--146.0 GHz);
and 1.3 mm (215.6--224.2 GHz).
The spectral resolution (channel width) was
$\Delta_{ch} = 200$ kHz, or 0.27 and 0.64 km s$^{-1}$ 
in the velocity scale
at the centers of the respective  edge bands at 1.3 mm and 3 mm.

The original paper \cite{Zhao} presents spectral observations of 148 massive molecular clouds (star-forming regions) located at various galactocentric distances in the range $0.14 < D_{GC} < 22.6$ kpc.
The distances to the selected objects are known with high precision thanks to measured trigonometric parallaxes and maser proper motions published in \cite{Reid14, Reid19}.
The aim of the study \cite{Zhao} was to search for methanol lines in thermal or maser emission (a complete list of 19 CH$_3$OH lines 
and the coordinates of the molecular clouds are provided 
in Tables 2 and 3 of \cite{Zhao}) and to measure the physical 
characteristics of star-forming regions. 

In addition to these important tasks, methanol emission lines in certain objects~-- characterized by simple, single-component Gaussian 
profiles~-- are noteworthy; even at a moderate signal-to-noise ratio 
($SNR \ga 10$), these lines allow for radial velocity measurements
 with sub-channel precision. 
Indeed,
the statistical error in the position of a single Gaussian is given
by the expression \cite{La82}:
\begin{equation}
\sigma_{stat} = 0.7\frac{\Delta_{ch}}{SNR}\sqrt{M}\, ,
\label{L1}
\end{equation}
where $M = FWHM/\Delta_{ch}$ is the full width at half maximum in units of channel width.

It follows from (\ref{L1}) that, for example, 
in the 1.3 mm band, where $\Delta_{ch} \simeq 0.27$ km s$^{-1}$,
the position of a methanol line with a typical width 
of $FWHM \sim 5-6$ km s$^{-1}$ and a signal-to-noise ratio 
of $SNR \gtrsim 10$ can be measured 
with an error of less than $\frac{1}{3}\Delta_{ch}$.
Moreover,
if in relation (\ref{E3}) both velocities $V_i$ and $V_j$
have comparable
errors and the difference in sensitivity coefficients
$|\Delta Q|$
is approximately equal to 3, then the statistical error of  \dmm\
in a single measurement
does not exceed $1.4\times10^{-7}$.
This example can be used to select the most suitable sources 
and spectral lines that allow for \dmm\ estimates at the level of 
a few units of $10^{-7}$.
 
To the above, one additional condition should be added: 
the methanol lines must have close frequencies and fall within 
the same spectral band during simultaneous exposures in order 
to eliminate systematic errors associated with frequency scale 
calibration and radio telescope pointing errors.

The molecular clouds from the study by \cite{Zhao}
in which CH$_3$OH lines with the required
characteristics are observed are listed
in Table~\ref{T1}.
The first column of this table lists the names of the objects; 
the coordinates~-- right ascension ($R.A.$) and declination 
($Dec.$)~-- are given in the second column, and the galactocentric distances ($D_{GC}$) in the third.
In total, the sample of objects covers the range of galactocentric distances $1.5 < D_{GC} < 13$ kpc.

%-------------------------Table 1
\begin{table}[h!]
\centering
\caption{
Coordinates of the selected objects and their galactocentric distances, $D_{GC}$, from \cite{Zhao}.
}
\label{T1}
\begin{tabular}{l c c c}
\hline
\multicolumn{1}{l}{Object} & $R.A.$(J2000) &$Dec.$ (J2000) &$D_{GC}$, kpc\\ 
\hline
G010.47+00.02&18 08 38.22 &--19 51 50.26&1.58\\
G010.62--00.38&18 10 28.56 & --19 55 48.73  &3.38\\
G029.95--00.01&18 46 03.74 &--02 39 22.32  &4.62\\
G037.42+01.51&18 54 14.34 & +04 41 39.64  &6.73\\
G078.88+00.70&20 29 24.82 & +40 11 19.59  &8.17\\
G092.67+03.07&21 09 21.73 & +52 22 37.08  &8.36\\
G135.27+02.79&02 43 28.56 &+62 57 08.38 &13.08\\
\hline
\end{tabular}
\end{table}

The selected $E$-methanol lines are listed in Table~\ref{T2},
which includes thermal lines with widths of
$v_{tur} \sim 2-3$ km s$^{-1}$
(here $v_{tur} = FWHM/2.255$ is the
root-mean-square turbulent
velocity)\footnote{Maser lines from the survey by \cite{Zhao}
have widths of $v_{tur} < 1$ km s$^{-1}$.}.
High Mach numbers ($>4$) indicate, as well, the turbulent nature 
of the motions in the objects from the survey by Ref.~\cite{Zhao}.
Given that the kinetic temperatures in dense star-forming regions 
are $T_{kin} \la 100$ K and the thermal velocities of CH$_3$OH 
molecules are $v_{th} \la 0.2$ km s$^{-1}$, the measured molecular line widths can be interpreted as the result of the convolution of the local  absorption coefficient with the large-scale velocity field within the region restricted by the telescope aperture.
At the same time, a large number of turbulent cells
(differing in linear dimensions,
gas density, temperature, and velocity)
results~-- through random mixing~--
in a normal distribution of velocity vector projections
onto the coordinate axes in the case of isotropic
turbulence; that is, in a symmetric, Gaussian-like
profile of a single line, provided there is no
blending with other lines.

Such symmetric lines must be centered at the origin of the comoving reference frame.
Nevertheless, possible hidden blends of small amplitude (or instrumental defects) can affect the measured position of the line center, shifting it slightly to one side or the other while leaving the line profile shape virtually indistinguishable from a Gaussian at moderate $SNR$ values.
Such random shifts of the line center (so-called ``Doppler noise'') 
can arise during estimates of \dmm\ using formula (\ref{E3}).
From this point of view, the difference between two radial velocities 
$\Delta V = V_i - V_j$ in (\ref{E3}) can be represented as the sum of 
two components \cite{L10}:
\begin{equation}
\Delta V = \Delta V_\mu + \Delta V_{dop}\, ,
\label{L1a}
\end{equation}
where $\Delta V_\mu$ is the shift due to the variation of $\mu$, and
$\Delta V_{dop}$ is the shift caused by Doppler noise.

The signal $\Delta V_\mu$ can be estimated statistically by 
calculating the sample mean:
\begin{equation}
\langle \Delta V \rangle = \langle \Delta V_\mu \rangle \, ,
\label{L1b}
\end{equation}
since $ \langle \Delta V_{dop} \rangle = 0$ \kms.
In this case, the signal dispersion is equal to
\begin{equation}
Var(\Delta V) = Var(\Delta V_\mu ) + Var(\Delta V_{dop})\, .
\label{L1c}
\end{equation}
As for the Doppler noise $Var(\Delta V_{dop})$, it can be minimized through the appropriate choice of molecular lines and the specific 
selection of molecular clouds.

In the same vein, the requirement that the kinematic emission 
profile be thermal rules out any potential shift of the line center 
(relative to the laboratory value) 
which is characteristic of maser lines.
First, the sizes of maser spots are many orders of magnitude smaller 
than the area covered by the telescope aperture, 
and multiple spatially separated maser spots are often observed 
within a single object~-- features that can be resolved only through interferometric observations with high angular resolution.
 Secondly, the methanol molecule possesses a hyperfine structure,
 and the maser emission is dominated by only one hyperfine component 
\cite{Lank}; consequently, accurately determining the line center 
becomes difficult in this case.
For example, the hyperfine splitting of torsional-rotational transitions 
in methanol at 25 GHz (ground torsional state, $v_t = 0$) is approximately 0.2 km~s$^{-1}$ \cite{GL, VLe24}.

It is also necessary to know the uncertainties of the laboratory transition frequencies in order to account for their contribution to the radial velocity error $\sigma_v$, which, in the general case, arises from two components:
 \begin{equation}
\sigma_v = \sqrt{\sigma_{stat}^2 + \sigma_{sys}^2},
\label{E5}
\end{equation}
where $\sigma_{stat}$ is the error of the spectral line position measurement and $\sigma_{sys}$ is the  laboratory error.

Ultimately, three methanol transitions listed in Table \ref{T2} satisfied our requirements.
All of them lie in the high-frequency 1.3 mm range
(where the spectral resolution is maximal,
$\Delta_{ch} \simeq 0.27$ \kms),
are characterized by a difference in sensitivity coefficients
$|\Delta Q| \geq 1.8$, and exhibit symmetric profiles
described by a single Gaussian (see Fig. 1 in \cite{Zhao}).

The characteristics of the selected transitions are as follows: the quantum numbers~-- $J$ (total angular momentum) and its projection
 $K$~-- for the upper ($u$) and lower ($\ell$) levels are given in the first column of Table~\ref{T2}; the second column lists the measured laboratory frequency from \cite{Xu97}; the energies of the upper levels ($E_u$) are indicated in the third column; and the sensitivity coefficient $Q$ from \cite{J11} is in the last column (note that in \cite{J11}, the sensitivity coefficient is $K_{\mu} = -Q$).
In the fourth and fifth columns, we present the critical densities
$n^{(50)}_{cr}$ and $n^{(100)}_{cr}$, calculated for
two kinetic temperatures, 50~K and 100~K, in the optically thin
case:
\begin{equation}
n_{cr} = \frac{A_{u\ell}}{\gamma_{u\ell}(T_{kin})}\, .
\label{L2}
\end{equation}
The values ​​of the Einstein coefficients $A_{u\ell}$ and the collisional 
de-excitation coefficients $\gamma_{u\ell}$ for collisions with H$_2$ molecules are taken from the Leiden molecular data 
database\footnote{http://home.strw.leidenuniv.nl/$\sim$moldata/ }.
 
 It is generally assumed that if the gas density is below the critical value ($n < n_{\text{cr}}$), the radiation at a given frequency is weak.
Otherwise ($n > n_{cr}$), the line luminosity is high and the upper level becomes thermalized, i.e., its excitation temperature approaches the kinetic temperature of the gas. 
However, these estimates are approximate, 
since they do not take into account the effects of multiple scattering 
in the line within the cloud.
The role of photon-trapping processes increases with optical depth, leading to values ​​of $n_{cr}$ derived from formula (\ref{L2})  that are several times too high  \cite{Shir}.
Thus, it can be expected that all three $E$-methanol lines from Table~\ref{T2} trace the same gas with a density of $n \sim 10^7$ cm$^{-3}$.

It should be noted that Table~\ref{T2} lists the theoretical frequency 
for the 220.078 GHz line, calculated using quantum-mechanical 
methods in \cite{Xu97, Xu2}, because this transition has not been observed under laboratory conditions.
However, the error for this frequency is taken as 50 kHz, since this value is characteristic of laboratory measurements in this frequency range (see \cite{Xu97}).
Furthermore, this error    
is consistent with the spread   
of theoretical values ​​for the frequency of this transition   
from the cited works:   
$220078.490\pm 0.013$ MHz \cite{Xu97}   
and $220078.561\pm0.008$ MHz \cite{Xu2}.

%-------------------------Table 2
\begin{table}[h!]
\centering
\caption{
Selected $E$-methanol (CH$_3$OH) lines from \cite{Zhao}. Laboratory frequencies are taken from \cite{Xu97}, and sensitivity coefficients $Q$ from \cite{J11}. 
The $1\sigma$ uncertainties in the last digits are given in parentheses.
}
\label{T2}
\begin{tabular}{lccccc}
\hline\\[-8pt]
\multicolumn{1}{l}{Transition} & Lab. frequency & $E_u/k$ &
$n^{(50)}_{cr}$ & $n^{(100)}_{cr}$ &  $Q$\\ 
\multicolumn{1}{l}{$J_{K_u} \to J_{K_\ell}$}& $f$, MHz &K& 
cm$^{-3}$ & cm$^{-3}$ & \\
\hline
{$5_{1} \to 4_{2}E$}& 216945.600(50)& 47.98&$4\times10^6$& $7\times10^6$&+3.2\\
{$4_{2} \to 3_{1}E$}& 218440.050(50)& 37.57&$4\times10^7$&$8\times10^7$&--1.2\\
{$8_{0} \to 7_{1}E$}& 220078.561(50)$^{\ast}$& 88.72&$2\times10^7$&$3\times10^7$&+0.6\\
\hline
\multicolumn{4}{l}{\footnotesize $^\ast$The theoretical 
frequency value with an error characteristic of 
}\\
\multicolumn{6}{l}{\footnotesize\,\,\, 
laboratory measurements in this frequency range.
}
\end{tabular}
\end{table}

%-----------------------Section-3

\section{Estimates of $\Delta\mu/\mu$}
\label{Sect-3}

The parameters of the observed CH$_3$OH lines for a sample of 7 objects are presented in Table~\ref{T3}.
The first column lists the names of the objects;
the second, the rounded laboratory frequencies in GHz;
the third, the measured radial velocities $V_{LSR}$;
the peak temperatures $T_{mb}$ and
signal-to-noise ratios $SNR$ are given in
the fourth and fifth columns, respectively (see Table~4 in \cite{Zhao}).
The radial velocity errors $\sigma_v$ shown in the third column were calculated using formula (\ref{E5}), taking into account the uncertainties in the laboratory frequencies (50 kHz, or 0.069 km~s$^{-1}$ in this case).
Thus, the indicated $\sigma_v$ values ​​ exceed the statistical uncertainties in determining line centers from Table~4 in \cite{Zhao}, where the positions of methanol spectral lines and their corresponding uncertainties were determined using the CLASS software 
package\footnote{Continuum 
Line Analysis Single-dish Software  of the Grenoble Image and Line
Data Analysis Software packages.}.

%-------------------------Table 3
\begin{table}[h!]
\centering
\caption{
Parameters of CH$_3$OH lines toward the sample objects from \cite{Zhao}.  The $1\sigma$ uncertainties in the last digits are 
given in parentheses.
}
\label{T3}
\begin{tabular}{l c c c c}
\hline
\multicolumn{1}{l}{{\bf Object}}  & $f$, & $V_{LSR}$, & 
$T_{mb}$, & $SNR$\\ 
&GHz&km s$^{-1}$& K &\\
\hline
{\bf G010.47+00.02}&218.440&66.31(8)&6.24&39\\
&220.078&66.59(12)&2.54&25\\[6pt]
{\bf G010.62--00.38}&216.945&--2.64(11)&0.87&22\\
&218.440&--2.69(7)&4.06&81\\[6pt]
{\bf G029.95--00.01}&216.945&97.86(13)&1.04&52\\
&218.440&97.95(7)&2.58&65\\
&220.078&98.12(11)&0.91&30\\[6pt]
{\bf G037.42+01.51}&216.945&43.95(17)&0.74&19\\
&218.440&43.93(9)&2.04&68\\[6pt]
{\bf G078.88+00.70}&216.945&--5.68(19)&0.42&21\\
&218.440&--5.63(8)&0.75&25\\[6pt]
{\bf G092.67+03.07}&216.945&--6.02(12)&0.18&16\\
&218.440&--5.85(7)&4.6&92\\
&220.078&--5.88(11)&0.91&10\\[6pt]
{\bf G135.27+02.79}&216.945&--71.73(14)&0.1&10\\
&218.440&--71.49(10)&0.36&36\\
\hline
\end{tabular}
\end{table}

The constraints on variations of $\mu$ and the pairs of lines from which they were derived are presented in Table~\ref{T4}. This table requires a number of explanations.
First, the complete set of methanol lines
from the 1.3 mm band (Table~\ref{T2}) was used only for
two sources: G029.95--00.01 and G092.67+03.07.
Secondly, the 220 GHz line was excluded from the calculations
in four cases: $(i)$ for G010.62--00.38
due to a shift of more than 3 channels relative to the other two
lines, which~-- despite having a maximum sensitivity coefficient
difference of $|\Delta Q| = 4.4$~-- agree within
$\frac{1}{5}\Delta_{ch}$;
$(ii)$ for G037.42+01.51 due to profile asymmetry;
$(iii)$ for G078.88+00.70 due to a low signal-to-noise ratio
($SNR = 7$); $(iv)$ for G135.27+02.79, where the line exhibits
a $\Pi$-shaped profile with a flat top.
Thirdly, the 216 GHz line in G010.47+00.02 
was not included in the \dmm\ estimate because 
it is blended with another
emission line clearly visible in the right wing of the
216 GHz line profile (see Fig.~1 in \cite{Zhao}).
 
%-------------------------Table 4
\begin{table}[h!]
\centering
\caption{Results of $\Delta\mu/\mu$ estimates.}
\label{T4}
\begin{tabular}{l l c}
\hline\\ [-8pt]
\multicolumn{1}{c}{{\bf Object}}  &
\multicolumn{1}{c}{Used lines} &$\Delta\mu/\mu \times 10^{-7}$\\ 
\hline
{\bf G010.47+00.02}&220\&218&$(-5.2 \pm 2.7)$\\[6pt]
{\bf G010.62--00.38}&216\&218&$(-0.4 \pm 1.0)$\\[6pt]
{\bf G029.95--00.01}$^a$ &218\&216 and 218\&220 
&$(0.2 \pm 1.1)$\\
&216\&220&$(3.3 \pm 2.2)$\\[6pt]
{\bf G037.42+01.51}&216\&218&$(-0.2 \pm 1.5)$\\[6pt]
{\bf G078.88+00.70}&216\&218&$(0.4 \pm 1.6)$\\[6pt]
{\bf G092.67+03.07}$^b$ &218\&216 and 218\&220
&$(1.2 \pm 1.1)$\\
&216\&220&$(1.8 \pm 2.1)$\\[6pt]
{\bf G135.27+02.79}&216\&218&$(1.8 \pm 1.3)$\\
\hline
\multicolumn{3}{l}{\footnotesize $^a$The weighted mean value of 
$\langle \Delta\mu/\mu\rangle$ for G029.95--00.01 is}\\
\multicolumn{3}{l}{\footnotesize\,\,\, 
equal to $(0.8\pm1.2)\times10^{-7}$.}\\
\multicolumn{3}{l}{\footnotesize $^b$The weighted mean value of 
$\langle \Delta\mu/\mu\rangle$ for
G092.67+03.07 is}\\
\multicolumn{3}{l}{\footnotesize\,\,\, 
equal to $(1.3\pm0.6)\times10^{-7}$.}
\end{tabular}
\end{table}

The calculations of \dmm\ for the remaining pairs of $E$-methanol lines from Table~\ref{T3} were performed using formula (\ref{E3}).
In this case, assuming statistically independent measurements of radial velocity differences ($V_i - V_j$), the error $\Delta\mu/\mu$ for a single pair of lines is given by the expression
\begin{equation}
\sigma_{ij} = 
\frac{\sqrt{\sigma^2_{v_i} + \sigma^2_{v_j}}}{c|Q_j - Q_i|}\, ,
\label{L3}
\end{equation}
in which the errors of the corresponding radial velocities, 
$\sigma_{v_i}$ and $\sigma_{v_j}$, were calculated 
using formula (\ref{E5}).

However, 
to calculate the mean errors $\sigma_{\langle\Delta\mu/\mu\rangle}$ 
for the objects G029.95--00.01 and G092.67+03.07~-- where 
radial velocity differences relative to the same 
218 GHz reference line are used~-- 
it is necessary to account for the statistical dependence between these 
differences, which leads to additional 
correlation corrections to formula (\ref{L3}).

In general, for a series of the measured differences
$V_1 - V_0, V_2 - V_0, V_3 - V_0 \ldots$, where all $V_i$ are
independent random variables and $V_0$ is a chosen 
reference velocity,
the correlation coefficient $\kappa_{ij}$
between the pair of differences $(V_i - V_0)$ and $(V_j - V_0)$
is calculated as \cite{VKL24, L10}
 \begin{equation}
\kappa_{ij} = \frac{1}{\sqrt{(1+\frac{\sigma_{v_i}^2}{\sigma_{v_0}^2})(1+\frac{\sigma_{v_j}^2}{\sigma_{v_0}^2})}} ,
\label{E6}
\end{equation}
where $\sigma_{v_0}$ is the error of the reference velocity 
(the 218 GHz line in our case), and $\sigma_{v_i}, \sigma_{v_j}$ 
are the velocity errors of the lines being compared (in our case, the 216 and 220 GHz lines, respectively).

The error of the weighted mean
$\langle \Delta\mu/\mu \rangle$ in this case will be equal to
(see, e.g., \cite{Coh})
 \begin{equation}
\sigma_{\langle \Delta\mu/\mu\rangle} = \frac{1}{W}\left[ 
\left(\frac{1}{\sigma^2_{i0}}+\frac{1}{\sigma^2_{j0}}\right) +
2\frac{\kappa_{ij}}{\sigma_{i0} \sigma_{j0}} 
\right]^{1/2},
\label{E7}
\end{equation}
where $\sigma_{i0}$ and $\sigma_{j0}$ are the errors of the \dmm\ estimates
based on the corresponding pairs ($V_i - V_0$) and ($V_j - V_0$)
(formula (\ref{L3})).
$W = \sum w_ k = \sum 1/\sigma^2_{k0}$ is the total weight, 
and the weighted mean $\langle \Delta\mu/\mu \rangle$ itself is calculated using the standard formula
 \begin{equation}
\langle \Delta\mu/\mu \rangle = \frac{1}{W} 
\sum w_k (\Delta\mu/\mu)_k\, .
\label{E8}
\end{equation}
Here, according to the accepted notations, $k = i$ (216 GHz line) and 
$k = j$ (220 GHz line), and
$$
\left(\frac{\Delta\mu}{\mu}\right)_i = 
\frac{V_i - V_0}{c(Q_0 - Q_i)},\,\,\,\,\,
\left(\frac{\Delta\mu}{\mu}\right)_j = 
\frac{V_j - V_0}{c(Q_0 - Q_j)}\, .
$$

The third pair~-- the 216 and 220 GHz lines in the spectra of
 G029.95--00.01 and G092.67+03.07~-- is formed by statistically independent radial velocities; 
 therefore, the error \dmm\ was calculated using formula (\ref{L3}).

Thus, a series of nine \dmm\ values ​​of unequal precision 
was obtained; they are listed in Table~\ref{T4}.
The procedure for processing series of data of unequal precision is described, for example, in \cite{A72}.
This procedure compares the weighted-mean error 
$\sigma_\xi$~-- obtained via the error propagation formula 
(referred to in \cite{A72} as the pre-adjustment error)~-- 
with the error $\sigma^\ast_\xi$ obtained after the adjustment 
of a series of measurements of unequal precision. 
It should be noted that $\sigma_\xi$ takes into account the mean errors of each element of such data sample but does not account for the deviations of these measurement elements from their weighted mean 
$\xi$, whereas $\sigma^\ast_\xi$ does account for such deviations.
In general, $\sigma_\xi \neq \sigma^\ast_\xi$, since both of these 
errors are random variables.  
 
 If $\sigma_\xi > \sigma^\ast_\xi$, the cause of the discrepancy 
 is purely random, and the arithmetic mean of these two
errors~-- $\frac{1}{2}(\sigma_\xi + \sigma^\ast_\xi)$~-- is 
adopted as the final value of the weighted mean error.
Otherwise, if $\sigma_\xi < \sigma^\ast_\xi$, then $\sigma^\ast_\xi$ is adopted as the final value of the weighted mean error,  
since the latter inequality is likely caused by systematic errors 
within the series of measurements.
 
Based on the described procedure, the following weighted mean 
is obtained for our dataset:
$\langle\Delta\mu/\mu\rangle_{disk} = (0.5\pm0.5)\times10^{-7}$,
which indicates the absence of statistically significant
variations in $\mu$ exceeding the level of $0.5\times10^{-7}$
within the range of galactocentric distances
$1.5 < D_{GC} < 13$ kpc.
Entirely different estimates are obtained,
as noted in the introduction,
near the Galactic center in the Sgr\,B2 object,
where, given comparable uncertainties, 
the mean value of $\Delta\mu/\mu$
differs significantly from zero:
$\langle\Delta\mu/\mu\rangle_{cent} = (-3.4\pm0.4)\times10^{-7}$.

It is worth separately discussing the measurements of $\Delta\mu/\mu$ 
in the molecular cloud G010.47+00.02, located relatively close to the Galactic center ($D_{GC} = 1.58$ kpc; Table \ref{T1}).
For this object, $\Delta\mu/\mu = (-5.2\pm2.7)\times 10^{-7}$,
which agrees within $1\sigma$ with measurements in Sgr\,B2.
However, the statistical significance of such a coincidence is low
due to the large error and the fact that \dmm\ is estimated
 on only two methanol lines.
In addition, the analysis utilizes the 220 GHz line, for which no laboratory measurements are available.
Nevertheless, the near-center molecular cloud G010.47+00.02 warrants further attention as a target for follow-up observations using a larger number of methanol lines to improve the accuracy of \dmm\ estimates.

%-----------------------------------Figure 1
\begin{figure}[h!]
%\vspace{-2cm}
\centering
\includegraphics[width=0.9\textwidth]{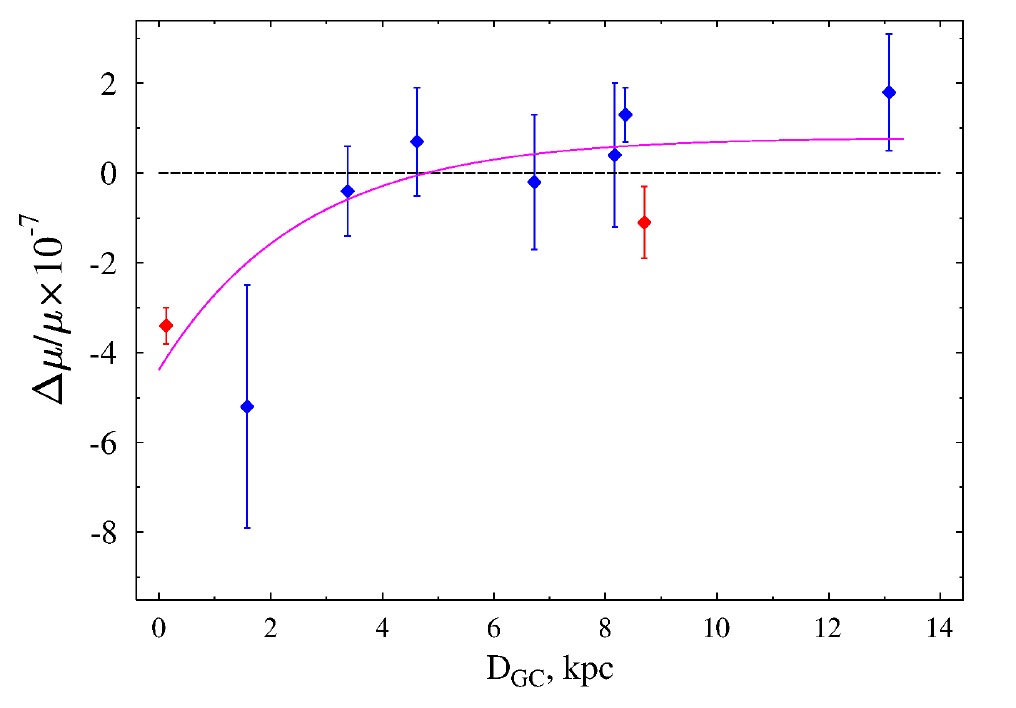}
%\vspace{-1.0cm}
\caption{\small
Dependence of $\Delta\mu/\mu$ on the galactocentric 
distance $D_{GC}$. 
Measurements of $\Delta\mu/\mu$ from \cite{VL26} are shown in red, while the results of the present work are in blue. The pink curve represents a fit to the exponential law $y = -5.2e^{-0.39x}+0.80$ 
($\chi^2_\nu = 1.46$). The horizontal dashed line marks the 
level $\Delta\mu/\mu = 0$.  
}
\label{F1}
\end{figure}

A graphical representation of the obtained data 
is shown in Fig.~\ref{F1}, which displays the measured 
values ​​of $\Delta\mu/\mu$ at various galactocentric distances.
Points corresponding to $\Delta\mu/\mu$ estimates derived from 
{\it thermal} methanol lines in Sgr\,B2 ($D_{GC} \simeq 0.1$ kpc) and Orion-KL ($D_{GC} = 8.7$ kpc) are marked in red, 
while the results of the present work are shown in blue.
The combined points are approximated (for illustrative purposes only)
by an exponential law:
$y = -5.2e^{-0.39x}+0.80$, $\chi^2_\nu = 1.46$
(with the number of degrees of freedom $\nu = 6$, the expected
standard deviation $\sigma(\chi^2_\nu)$ is equal to
$\sqrt{2/\nu} = 0.577$ \cite{Wu}).
The decline of the approximating curve at small values ​​of $D_{GC}$ and its near-zero values ​​far from the center ($D_{GC} > 3$ kpc) may indicate a differing influence of scalar fields on elementary particle masses in the Galactic center and at the periphery.
However, this statement requires further experimental verification.

%-------------------------------------Section 4
\section{Conclusion}
\label{Sect-4}

The main goal of this work is to estimate $\Delta\mu/\mu$ using thermal lines of $E$-methanol (CH$_3$OH) toward massive molecular clouds in the Galactic disk located at various galactocentric distances, including a cloud near the Galactic center.
Despite the fact that the Galactic center is characterized by a peaked rise in dark matter density (a ``cuspy halo'', \cite{NFW}), this region remains understudied in the context of searching for variations in fundamental physical constants.
Therefore, obtaining new data on hypothetical variations of $\mu$ across the entire Galactic disk is of undoubted interest.
We emphasize once again that the hypothesized connection between dark matter and the Higgs scalar field must be of a large-scale nature and affect the values ​​of fundamental physical constants independently of the local density of baryonic matter and/or the gravitational potential of individual objects, as considered, for example, in certain theoretical models \cite{MBS} – \cite{Brax}.

In conclusion, we highlight the main results of the work performed.
\begin{enumerate}
\item
Constraints have been set on $\mu$-variations in the Galaxy in the direction of seven massive star-forming regions from the survey by \cite{Zhao}:
 $\langle\Delta\mu/\mu\rangle_{disk} = (0.5\pm0.5)\times10^{-7}$.
 The selected objects cover practically the entire visible galactic disk
  ($1.5 < D_{GC} < 13$ kpc).
Taking into account all possible systematic effects, the calculated weighted mean $|\langle\Delta\mu/\mu\rangle_{disk}|$ does not exceed 
$5 \times 10^{-8}$, which is consistent with the upper limits for individual molecular clouds in the Galactic disk discussed in the introduction.
  
\item
An estimate of $\Delta\mu/\mu$ has been obtained for the molecular cloud G010.47+00.02, located close to the Galactic center
 ($D_{GC}$ = 1.58 kpc).
Although in this case 
$\Delta\mu/\mu = (-5.2 \pm 2.7) \times 10^{-7}$ 
is negative~-- as is the value for the object 
Sgr\,B2 ($D_{GC} \simeq 0.1$ kpc)~--
the large measurement uncertainty 
precludes drawing any statistically significant conclusion 
regarding the signal.
To verify the obtained result, it is necessary to conduct independent studies using a larger number of methanol transitions and more precise measurements of spectral line centers.

\item
The discussed procedure for selecting spectral lines and molecular 
clouds~-- suitable for precision measurements of $\mu$ variations in the Galaxy and for testing the hypothesized link between dark matter and the Higgs scalar field~-- can serve as a starting point for further research at higher levels of precision, specifically $10^{-8}$ and beyond.

\end{enumerate}

\subsection*{Funding}
This work was funded by the budgets of Electrotechnical University 
``LETI'' and the Ioffe Institute.
No additional grants were received for conducting or supervising this specific study.
  
\subsection*{Conflict of interest} 
The authors of this work declare that they have no conflicts 
of interest.

\subsection*{Acknowledgments} 
The authors would like to thank the anonymous reviewer
for his/her constructive comments.

\end{document}